\documentclass[prb,twocolumn,draft,amsmath,showpacs]{revtex4}
\usepackage{graphics}\input epsf
\begin{document}
\title {Giant phonon anomalies in the pseudo-gap phase of TiOCl}

\author {P. Lemmens$^1$, K.Y. Choi$^2$, G. Caimi$^3$, L. Degiorgi$^{3,4}$,
N.N. Kovaleva$^1$, A. Seidel$^5$, and F.C. Chou$^5$}

\address{$^1$Max Planck Institute for Solid State Research, D-70569
Stuttgart, Germany}

\address{$^2$2. Physikalisches Institut, RWTH
Aachen, D-52056 Aachen, Germany}

\address{$^3$Laboratorium f\"ur
Festk\"orperphysik, ETH Z\"urich, CH-8093 Z\"urich, Switzerland}

\address{$^4$Paul Scherrer Institute, CH-5232
Villigen, Switzerland}

\address{$^5$Center for
Materials Science and Engineering, MIT, Cambridge, MA 02139, USA}

\date{July, 22nd, 2003}

\begin{abstract}
We report infrared and Raman spectroscopy results of the spin-1/2 quantum
magnet TiOCl. Giant anomalies are found in the temperature dependence of the
phonon spectrum, which hint to unusual coupling of the electronic degrees of
freedom to the lattice. These anomalies develop over a broad temperature
interval, suggesting the presence of an extended fluctuation regime. This
defines a pseudo-gap phase, characterized by a local spin-gap. Below 100~K a
dimensionality cross-over leads to a dimerized ground state with a global
spin-gap of about 2$\Delta_{spin}\approx$~430~K.
\end{abstract}
 \maketitle

Recently a new class of spin-1/2 transition metal oxides has been identified
based on $\rm Ti^{3+}$ ion in a distorted octahedral coordination of oxygen
and/or chlorine \cite{axt97,isobe02,seidel03,isobe02b}. Such compounds
especially in two dimensions (2D) are candidates for exotic electronic
configurations \cite{beyon93} and for superconductivity based on dimer
fluctuations \cite{seidel03,seidel03b}. Despite different stoichiometry and
structural details their low-energy degrees of freedom are characterized by
spin-1/2 quantum magnetism with a singlet ground state as a result of a
phase transition. Some of these aspects are reminiscent of a spin-Peierls
instability \cite{bray83}. However, significant differences from such a
scenario show up as an extended fluctuation regime above the transition
temperature, an extremely large magnitude of the singlet-triplet excitation
spin-gap and pronounced phonon anomalies. Therefore, it is tempting to
assign part of the dynamics of these phenomena to large electronic energy
scales, associated with orbital degrees of freedom. This idea stems from the
orbital degeneracy of $\rm t_{2g}$ states of the $\rm Ti^{3+}$ ions ($\rm
3d^1$, s=1/2) in octahedral surrounding and from the fact that their orbital
ordering can be destabilized by quantum fluctuations
\cite{khaliullin00,brink02}. Furthermore, novel mechanisms, based on the
creation of a coherent spin-orbital structure
\cite{pati98,yamashita00,koleshuk01}, have been predicted in order to obtain
a spin-gap state and coupled spin-orbital modes. Since the lattice is
strongly coupled to the orbital state, pronounced phonon anomalies are
expected. Up to now for none of the discussed titanates
\cite{axt97,isobe02,seidel03,isobe02b} a combined study of the spin and
phonon excitation spectrum is available. This is partly related to the
problem of growing sufficiently large single crystals for spectroscopic
studies.

In this letter we present an infrared (IR) absorption and Raman scattering
study on high quality single crystals of insulating $\rm TiOCl$. We
demonstrate an unusual coupling of the electronic degrees of freedom to the
lattice which is manifested by pronounced phonon anomalies and an extended
fluctuation regime, defining a pseudo-gap phase with a characteristic
temperature $\rm T^*$$\approx$~135~K.

TiOCl is a 2D oxyhalogenide formed of $\rm Ti^{3+}O^{2-}$ bilayers,
separated by $\rm Cl^-$ bilayers. The basic $\rm TiCl_2O_4$ octahedra build
an edge-shared network in the ab-plane of the orthorhombic unit cell. Above
100~K the magnetic susceptibility $\rm \chi(T)$ of TiOCl is only weakly
temperature dependent and forms a broad maximum at $\rm T_{max}$~=~400~K.
This motivated the proposal of a RVB ground state for this system
\cite{beyon93}. However, $\rm \chi(T)$ can be fitted using a s=1/2
Heisenberg spin chain model with an AF exchange coupling constant J=660~K
(Ref.~\onlinecite{seidel03}). Furthermore, $\rm \chi(T)$ displays a sharp
drop and a kink at $\rm T_{c1}$~=~66~K and $\rm T_{c2}$=94~K (Ref.
\onlinecite{seidel03,imai03}), respectively, with a small anomaly in the
specific heat $\rm c_p$(T) (Ref.~\onlinecite{lee03}) at $\rm T_{c2}$ only.
The first order phase transition at $\rm T_{c1}$ also involves a static
structural component that leads to a doubling of the unit cell in b-axis
direction \cite{lee03}. The origin of the 1D exchange path has been
identified as a direct exchange of Ti $\rm d_{xy}$ states aligned along the
b-axis. From band structure calculations it is deduced that the other two
d$\rm _{xz}$ and d$\rm _{yz}$ orbitals remain degenerate and are shifted to
higher energies \cite{seidel03}. The comparably large antiferromagnetic
exchange is unlikely to be understood solely on the basis of direct
exchange. Additional superexchange pathes would involve oxygen sites that
connect Ti-sites of the upper and lower bilayer and contribute to a possible
2D character of the spin system. As we will see in the following, such a
scenario makes TiOCl an unique system to study the effect of coupled
spin/lattice fluctuations on the basis of a complex exchange topology.

The first microscopic information about this interplay came from NMR/NQR
experiments \cite{imai03} on $\rm ^{35}Cl$ and $\rm ^{47,49}Ti$. The
relaxation rate of $\rm ^{35}Cl$ sites indicates dynamic lattice distortions
with a high temperature onset at $\rm \approx$~200~K. For the $\rm
^{47,49}Ti$ sites, $\rm (TT_1)^{-1}$, which probes the spin degrees of
freedom, reaches a maximum at T$^{*}$=135~K. The temperature dependence of
$\rm (TT_1)^{-1}$ implies a pseudo-gap phase in the homogeneous state of the
spin system with an estimated pseudo-gap $\rm \Delta_{fluct}$$\approx$430~K
(Ref.~\onlinecite{imai03}). This is extraordinarily large when compared to
the transition temperatures, leading to $\rm 2\Delta_{fluct}$/$\rm
k_BT_{c}$=9.1 and 13, for $\rm T_{c2}$ and $\rm T_{c1}$, respectively. These
ratios are not consistent with a spin-Peierls mechanism for this gap
formation \cite{bray83}.

Optical reflectivity $R(\omega)$ and Raman scattering experiments have been
performed with light polarization within the ab-plane. The Kramers-Kronig
transformation of $R(\omega)$ allows us to evaluate the real part
$\sigma_{1}(\omega)$ of the optical conductivity. Details pertaining to the
optical experiments can be found elsewhere \cite{wooten,lemmens-rev}. Due to
the $\rm D_{2h}$ point group symmetry of the atoms in TiOCl, optical
reflectivity and Raman spectroscopy complement each other. With light
polarization within the ab-plane they exclusively probe in-plane and
out-of-plane (c-axis) displacements, respectively.

\begin{figure}[t]
     \begin{center}
      \leavevmode
      \epsfxsize=7cm \epsfbox{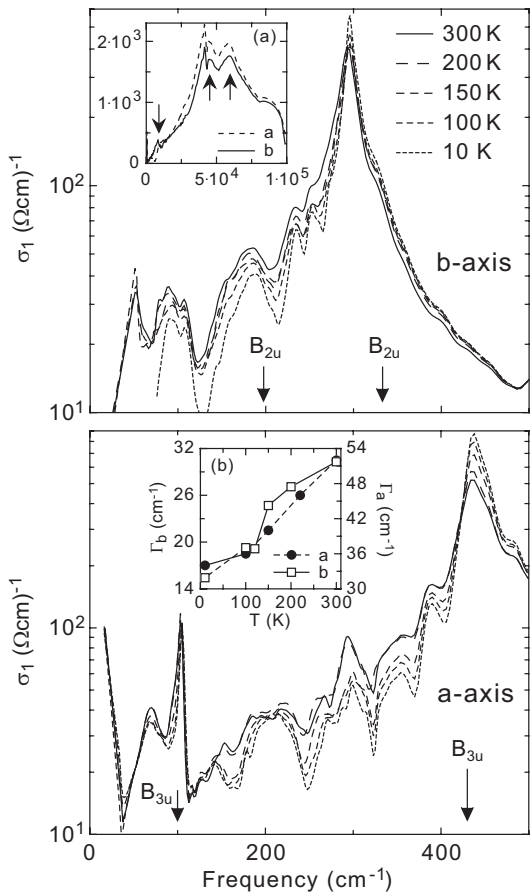}
       \caption{Real part $\sigma_{1}(\omega)$ of the optical conductivity
       of $\rm TiOCl$ as a function of
       temperature along the b-axis and the a-axis. The
       inset (a) shows $\sigma_{1}(\omega)$ up to the ultra-violet at 300~K.
       The inset (b) displays the
       temperature dependence of the width ($\Gamma$) of the phonon modes
       at 294 and 438~$\rm cm^{-1}$, respectively.  }
\label{sigma1}
\end{center}
\end{figure}


IR-spectroscopy results (Fig.~1) in the far infrared spectral range show a
strong anisotropy of the optical response within the ab-plane, indicative of
the low dimensionality of $\rm TiOCl$. The inset (a) displays the absorption
spectrum up to the ultraviolet spectral range at 300~K, and emphasizes the
energy interval dominated by the electronic interband transitions. The
observed peak at $\rm \sim 10^{4} ~cm^{-1}$ (1 eV) as well as maxima between
3.2x10$\rm ^{4}$ and 5.6x10$\rm ^{4} ~cm^{-1}$ (4 and 7~eV) (down-arrow and
up-arrows in inset (a) of Fig.~1) are interpreted as the predicted splitting
of the t$_{2g}$ states and the interband transitions between the O and Cl
p-levels and the Ti d-levels, respectively \cite{seidel03}. The
$\sigma_{1}(\omega)$ spectra above 2x10$\rm ^{4}$ $cm^{-1}$ turn out to be
polarization independent.

In the far-infrared (main panels of Fig.~1) $\sigma_{1}(\omega)$ is
dominated by strong peaks at 294~$\rm cm^{-1}$ and at 438 $\rm cm^{-1}$
along the chain b-axis and the transverse a-axis, respectively. In addition,
broad and less intense modes are detected along both polarization
directions. Considering the space group $Pmmn (59)$ of $\rm TiOCl$ at room
temperature and light polarization in the ab-plane, two $\rm B_{3u}$ modes
polarized along the a-axis and two $\rm B_{2u}$ modes along the b-axis are
expected as infrared active phonons. Our calculations predict the $\rm
B_{3u}$ phonons at 100 and 431 $\rm cm^{-1}$ and the $\rm B_{2u}$ ones at
198 and 333 cm$^{-1}$  (see down-arrows in Fig.~1). The two high frequency
ones can be identified with the most pronounced features in the spectra
\cite{caimi03}. The larger number of broad and less intense modes are
attributed to a lower symmetry than assumed so far. The anisotropy of the
optical spectra allows us to exclude twinning or leakage effects of the
polarizer.

We note that with decreasing temperature the two strongest modes at 294~$\rm
cm^{-1}$ and at 438~$\rm cm^{-1}$ display a pronounced narrowing of their
linewidth $\Gamma$. The temperature dependence of $\Gamma$ (inset (b) of
Fig.~1) shows a 49\% and 35\% decrease between 300 and 10~K along the b- and
a-axis, respectively. This strong renormalization for temperatures higher
than $\rm T_{c2}$ is in agreement with the NMR \cite{imai03} and ESR results
\cite{kataev}. Furthermore, the spectral weight at low frequencies tends to
decrease below 200~K (Fig.~1). The total spectral weight, obtained by
integrating $\sigma_{1}(\omega)$ is, however, conserved and it is fully
recovered by $\rm 10^{4}~cm^{-1}$. A detailed analysis of the IR-phonon
spectra and their temperature dependence will be presented elsewhere
\cite{caimi03}.

\begin{figure}[t]
     \begin{center}
      \leavevmode
      \epsfxsize=7cm \epsfbox {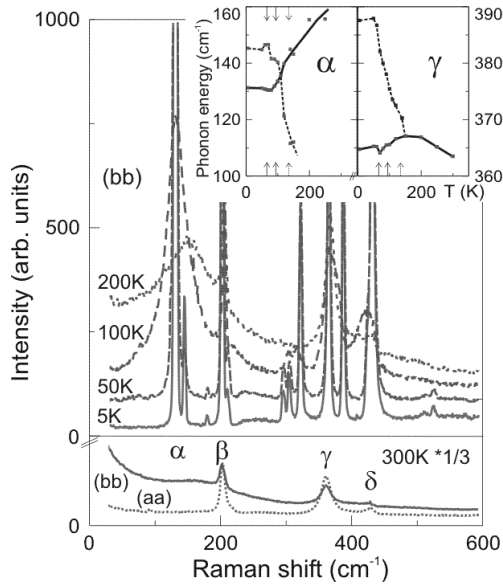}
       \caption{Raman scattering intensity as a function of temperature in (bb)
       polarization and with an offset for clarity.
       The lower inset compares (bb) and (aa)
       polarization at T=300~K with the intensity reduced by a factor 1/3.
       The four most important modes are denoted by $\alpha$ to $\delta$.
       The upper inset shows mode energies on heating the sample.
       The arrows mark $\rm T_{c1}$, $\rm T_{c2}$ and $\rm T^*$.
       Modes connected by dashed lines appear for T$<$$\rm T^*$ and
       have a smaller intensity.   }
\label{raman1}
\end{center}
\end{figure}

For Raman scattering with the light polarized within the ab-plane three
phonon modes are expected for the $Pmmn (59)$ space group. These A$_{g}$
modes have only displacements along the c-axis of the unit cell and are
predicted at 247, 333 and 431~$\rm cm^{-1}$ within the shell model
calculation. The Raman spectra with (aa) and (bb) polarization, shown in the
lower inset of Fig.~2, indeed display three modes at 203, 365 and 430~$\rm
cm^{-1}$, denoted by $\beta$, $\gamma$ and $\delta$, respectively. However,
the response in (bb) polarization, parallel to the chain direction of the
$\rm t_{2g}$ orbitals, is dominated by a very broad scattering continuum
with a maximum at about 160~$\rm cm^{-1}$, denoted by $\alpha$. With
decreasing temperature its linewidth strongly decreases and the maximum of
the $\alpha$-mode softens down to 130~$\rm cm^{-1}$, i.e. by $\approx$20\%.
This very large softening occurs in the fluctuation regime between 200~K and
$\rm T_{c1}$. The energy of this mode is moreover comparable to the
characteristic temperature T$^{*}$ of the pseudo-gap phase determined by
NMR. Also the other modes change appreciably by splitting into several sharp
components as shown in the inset. Finally, for T$<$$\rm T_{c1}$ all
transition-induced modes have a sharp and comparable linewidth and no more
anomalies are observed (see Fig.~2).


\begin{figure}[t]
     \begin{center}
      \leavevmode
      \epsfxsize=7cm \epsfbox {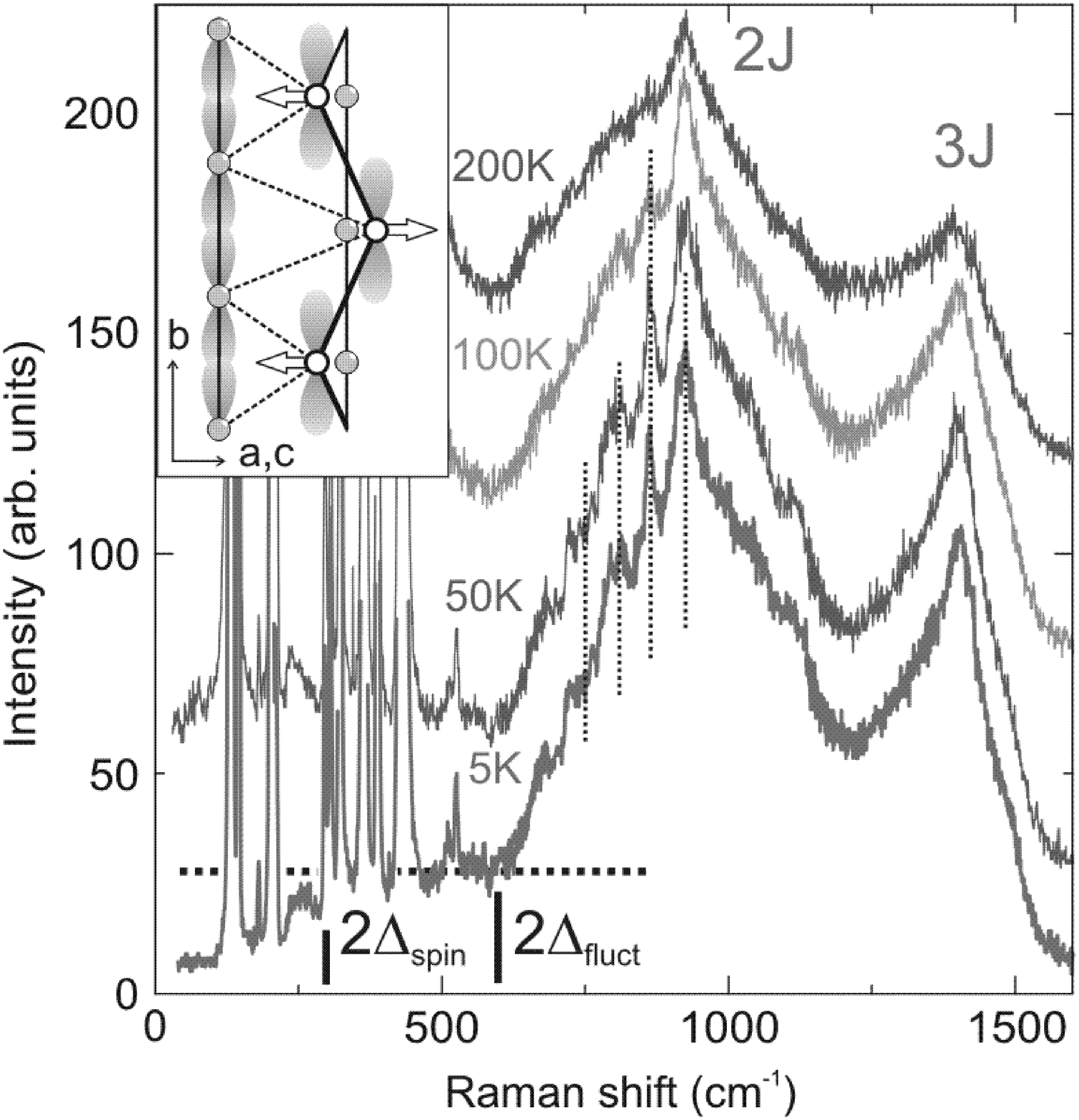}
       \caption{High energy Raman scattering intensity in (bb)
polarization with maxima at 920 (2J) and 1400~$\rm cm^{-1}$ (3J) and an
offset for clarity. The two gaps 2$\rm \Delta_{spin}$ and 2$\rm
\Delta_{fluct}$ are assigned by bars. A modulation of the scattering
intensity is marked by dashed lines. The inset shows a $\rm
Ti^{3+}$-displacement assigned to the $\alpha$-phonon mode. For clarity only
the orbitals d$_{xy}$ along the chain are shown. The dashed lines represent
d$\rm _{xz}$ and d$\rm _{yz}$ orbitals.} \label{raman2}
\end{center}
\end{figure}

In the low-energy range and for T$<$$\rm T_{c1}$, a change of a continuum of
scattering is visible in Fig.~3. This gradual depletion has an onset
frequency of 300~$\rm cm^{-1}\approx$430~K. Since similar effects have been
observed in $\rm NaV_2O_5$ and $\rm Sr_{14}Cu_{24}O_{41}$ at the doubled
spin-gap of the systems, see, e.g. Ref.~\onlinecite{lemmens-rev}, we also
attribute this onset in TiOCl to 2$\rm \Delta_{spin}$=300~$\rm cm^{-1}$. The
spin-gap 2$\rm \Delta_{spin}$ is about a factor of two smaller than the
pseudo-gap 2$\rm \Delta_{fluct}$ determined by NMR \cite{imai03} and leads
to more reasonable gap ratios of $\rm 2\Delta_{spin}$/$\rm k_BT_{c}$=4.6 and
6.7, for $\rm T_{c2}$ and $\rm T_{c1}$, respectively. The larger ratio
compared to the mean-field result ($\rm 2\Delta$/$\rm k_BT_{c}$=3.52) might
be attributed to competing exchange paths or electronic degrees of freedom.
Also in, $\rm NaTiSi_2O_6$, a related 1D titanate, orbital ordering at $\rm
T_c$=210~K leads to a large spin-gap with $\rm 2\Delta_{spin}$/$\rm
k_BT_{c}$=4.8 (Ref.~\onlinecite{isobe02}).


Figure 3 shows the Raman scattering in the energy range comparable to the
exchange coupling constant J. With polarization parallel to the chain axis
(bb) two maxima, a symmetric and an asymmetric one, are observed at
frequencies corresponding to 2J and 3J. The first maximum (2J) resembles the
two-magnon continuum of the spin tetrahedra system $\rm Cu_2Te_2O_5Br_2$,
which is in the proximity to a quantum critical point
\cite{lemmens01,gros03}. The second maximum (3J) is not expected within a
simple spin Hamiltonian and a single exchange path \cite{lemmens-rev} and
might be related to the proposed superexchange path. Its higher energy is
consistent with a larger coordination number of the involved magnetic sites
in 2D. A better understanding of the electronic structure of TiOCl would
allow to further analyze the high energy scattering. There is no pronounced
temperature dependence of the high energy scattering with the exception of a
moderate reduction in intensity. Therefore, a dramatic change of the local
hopping pathes and the involved orbital states is not supported in the
investigated temperature regime. The left, low-energy edge of the
pyramid-like scattering contribution can be interpreted as a second onset of
local spin excitations at 600~$\rm cm^{-1}$$\approx$860~K. This two-particle
excitation gap is identical in energy to the pseudo-gap 2$\rm
\Delta_{fluct}$ identified in NMR \cite{imai03}. It also exists for
temperatures T$>$$\rm T_{c2}$.

For Raman shifts $\Delta\omega$$\leq$2J a modulation of the scattering
intensity is observed with a characteristic energy of 60-70~$\rm cm^{-1}$.
This value corresponds quite well to the energy separation between the
$\beta$-mode and the $\alpha$-mode. Such an effect implies an effective
modulation of the exchange coupling by the two modes and a large density of
states over a considerable phase space. A somewhat comparable signature has
been found in $\rm Sr_{14}Cu_{24}O_{41}$ due to a charge ordering-induced
modulation of the exchange in the rungs of the spin ladders
\cite{schmidt03}. Furthermore, we observe that the $\alpha$ mode for
T$<$$\rm T_{c1}$ has the same appearance and linewidth as the other phonons,
while the intensity of magnetic scattering is at least one order of
magnitude smaller.

From our point of view, the fluctuation regime is mainly characterized by
the large anomalies of the $\alpha$-mode and its peculiar coupling to the
spin excitation spectrum. All observations point to a phonon as the origin
of the $\alpha$-mode. As the symmetry-allowed modes ($\beta$, $\gamma$,
$\delta$) have all been observed and the large intensity of the
$\alpha$-mode is not consistent with a smaller, local symmetry breaking, we
attribute the $\alpha$-mode to the Brillouin zone boundary. Certainly, due
to its close proximity in energy, the $\beta$-mode is its related zone
center Raman-allowed phonon mode. Lattice shell model calculations show that
the respective displacement is a pure c-axis in-phase Ti-Cl mode. A
projection of the corresponding zone-boundary displacement onto two adjacent
Ti chains is given in the inset of Fig.~3. The displacement leads to an
alternating deflection of the Ti sites out of the b-axis chain. The effect
of such a displacement on the electronic states can be quite drastic. The
higher lying $\rm t_{2g}$ $\rm d_{xz}$ and $\rm d_{yz}$ states, that mediate
the exchange perpendicular to the chains, are admixed to the ground state
and the respective hopping matrix elements should be enhanced
\cite{valenti03}. Further arguments that point into this direction are the
large magnitude of the exchange coupling along the chain axis with respect
to the phonon frequency and that J is only one order of magnitude smaller
than the splitting of the $\rm t_{2g}$ levels \cite{seidel03}. A coupling of
low and high energy scales via the quasi-degenerate orbitals of the $\rm
Ti^{3+}$ sites is a peculiarity of the present and the earlier mentioned
titanates. The pronounced softening and large scattering intensity of the
$\alpha$-mode in the pseudo-gap regime is therefore a direct fingerprint of
the coupled spin-lattice fluctuations. As the softening is strongly
nonlinear the change of population number are large and not comparable with
usual effects of anharmonicity.

Finally, we address the peculiar ordering phenomena with the three
characteristic temperatures $\rm T_{c1}$, $\rm T_{c2}$ and T$^{*}$ that in
our opinion reflect the interplay of thermal and quantum fluctuations. While
with decreasing temperatures (T$\leq$T$^{*}$) the coherence length of the
structural distortion slowly increases, the magnetic correlations cross-over
from 2D to 1D due to a change of the $\rm t_{2g}$ orbital admixture. The
energy gain for T$\leq$$\rm T_{c2}$ is mainly related to the spin system.
Therefore the related anomaly in the specific heat is small \cite{lee03} and
in conventional x-ray scattering no sign of a coherent structural distortion
can be found \cite{kataev}. The structural distortions become long range
only for T$\leq$$\rm T_{c1}$ as consequence of an order-disorder transition
due to the significant spin-lattice coupling \cite{lee03}. Below this
temperature the system shows a more conventional spin-Peierls-like behavior
with a global spin-gap 2$\rm \Delta_{spin}$. Consequently, the spin-gap
2$\rm \Delta_{fluct}$ is the smallest energy for a local double-spin-flip in
the short-range-order distorted (pseudo-gap) phase. Therefore, it is not
related to the transitions at $\rm T_{c1}$ and $\rm T_{c2}$.

To conclude, an extended fluctuation regime has been investigated in the
$\rm Ti^{3+}$ bilayer system TiOCl using IR and Raman spectroscopy. It is
proposed that a dimensionality crossover of the spin subsystem is induced by
the population of a soft phonon with an energy comparable to the fluctuation
scale T$^{*}$ of the pseudo-gap phase. We anticipate that to further
substantiate our scenario neutron scattering investigations or inelastic
x-ray scattering would be very important together with further theoretical
modelling.

\acknowledgments The authors acknowledge fruitful discussions with C. Gros,
R. Valenti, T. Saha-Dasgupta, R. Kremer, A. Perucchi, V. Gnezdilov, B.
Batlogg, H. Thomas, B. Keimer, and P.A. Lee. This work was supported
primarily by the MRSEC Program of the National Science Foundation under
award number DMR 02-13282, DFG SPP1073, NATO PST.CLG.9777766, INTAS 01-278,
and the Swiss National Foundation for the Scientific Research.

\end{document}